# Wireless Integrated Authenticated Communication System (WIA-Comm)

Amith N Bharadwaj
*Dept. of E&C Engineering*
*The National Institute of Engineering*
Mysuru, Karnataka, India
2020ec_amithnbharadwaj_a@nie.ac.in

G Adarsh
*Dept. of E&C Engineering*
*The National Institute of Engineering*
Mysuru, Karnataka, India
2020ec_gadarsh_a@nie.ac.in

Gurusatwik Bhatta N
*Dept. of E&C Engineering*
*The National Institute of Engineering*
Mysuru, Karnataka, India
2020ec_gurusatwikbhattan_a@nie.ac.in

Karan K
*Dept. of E& C Engineering*
*The National Institute of Engineering*
Mysuru, Karnataka, India
2020ec_karank_a@nie.ac.in

Dr. Vijay B T
*Dept. of E&C Engineering*
*The National Institute of Engineering*
Mysuru, Karnataka, India
vijaybt@nie.ac.in

*Abstract*—The exponential increase in the number of devices connected to the internet globally has led to the requirement for the introduction of better and improved security measures for maintaining data integrity. The development of a wireless and authenticated communication system is required to overcome the safety threats and illegal access to the application system/data. The WIA-Comm System is the one that provides a bridge to control the devices at the application side. It has been designed to provide security by giving control rights only to the device whose MAC (physical) address has already been registered, so only authorized users can control the system. LoRa WAN technology has been used for wireless communication and Arduino IDE to develop the code for the required functionality.

*Keywords—ESP8266, LoRa (Long Range) Communication, Arduino UNO, Authentication,*

## I. INTRODUCTION

As we know that, with the exponential increase in the number of devices and also devices connected to the internet are constantly increasing at a large scale, there is a vital requirement of proper security measures to protect data and systems. To achieve this, we need to come up with appropriate authentication and authorization measures for the scope of maintaining data security and integrity. The use of wireless communication systems also enables users to remotely access the data and the resources and also for securely controlling the systems. When the features of wireless communication and network security are integrated, we can come up with an effective and secure way to control or access the data remotely.

The use of proper hardware and technologies is essential in designing a system that provides all the required specifications. In this WIA-Comm system, we have used LoRa technology for communication, and Arduino interfaced with the LoRa board as the microcontroller.

For the application system i.e. the system to be controlled, a servo motor and an LED have been interfaced to mimic a motor and the lights in a factory. The microcontroller used is the LoRa interfaced with the Arduino board from enthutech. An ESP8266 board has been used to determine the MAC address of the device that is trying to enter the system to control the application system. We have also incorporated some additional functionality like an alert message after 3 illegal attempts

## II. LITERATURE SURVEY

Access control in cybersecurity is a critical aspect for ensuring the security and integrity of networks and systems. Traditional methods of authentication, such as passwords and cryptographic keys, have inherent vulnerabilities that can be exploited by malicious actors. In response to these challenges, researchers have explored alternative approaches, including MAC address-based authentication systems.

Tiwari et al. (2018) proposed a Secured MAC Address-Based Login System, presenting a novel approach to access control. Their study highlights the potential of MAC addresses as unique identifiers for authentication purposes. By leveraging MAC addresses, the system offers a more secure and reliable means of access control compared to traditional password-based methods.

Hong (2008) investigated Secure MAC address-based Authentication on X.509 v3 Certificate in Group Communication. This study explored the integration of MAC address-based authentication with X.509 v3 certificates, a widely used standard for public key infrastructure (PKI). By combining these technologies, the system provides enhanced security for group communication scenarios, mitigating the risk of unauthorized access and data breaches.

In a recent study, Bairwa and Joshi (2021) proposed Mutual authentication of nodes using session token with fingerprint and MAC address validation. This research introduces a comprehensive authentication mechanism that incorporates session tokens, fingerprint validation, and MAC address verification. The system aims to strengthen access control by ensuring mutual authentication between nodes, thereby enhancing the overall security of networked environments.

Overall, these studies underscore the importance of MAC address-based authentication as a viable alternative to traditional access control methods.

### III. COMPONENTS

#### A. NodeMCU ESP8266

The NodeMCU ESP8266 is an open-source development board and firmware kit that can be used to build IoT products. It has firmware that runs on the ESP8266 Wi-Fi SoC from Espressif Systems and hardware based on the ESP-12 module. The NodeMCU ESP8266 is powered by the ESP8266EX chipset, which includes a 32-bit Tensilica microcontroller. It also has built-in Wi-Fi connectivity. It provides a versatile and cost-effective approach to connect devices to the internet.

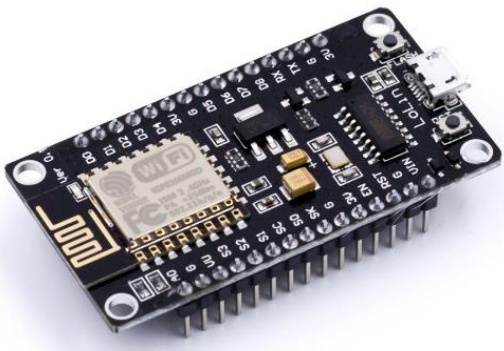

Fig 1. NodeMCU ESP8266

The LoLin NodeMCU V3 microcontroller has been selected and all the functionalities are coded and dumped using the very popular Arduino IDE. The NodeMCU ESP8266 is compatible with Arduino IDE and micro Python, making prototyping faster. It can be programmed with the LUA programming language or Arduino IDE.

#### B. LoRa Module

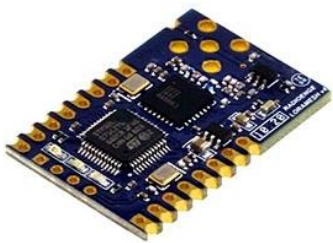

Fig 2. LoRaWAN

LoRa devices and the LoRa WAN standard offer compelling features for IoT applications including long-range, low power consumption, and secure data transmission. The technology has a longer range than cellular networks and is used by public, private, and hybrid networks. In rural locations, LoRa can connect devices up to 30 miles apart. It can even penetrate deep indoor environments and crowded metropolitan areas.

#### C. DC Motor

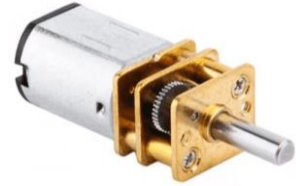

Fig 3. Stepper Motor

The N20 shaft motor, also known as DC mini metal gear motor is a small but powerful motor widely used in various electronics and robotics projects. It features a compact cylindrical design, typically around 12 mm in diameter and 10-15 mm in length, which makes it ideal for applications where space is limited. Operating at low voltages, usually between 3V and 12V, the N20 motor is known for its versatility in providing different speeds and torque levels, thanks to its built-in metal gearbox. This gearbox allows the motor to deliver higher torque, making it suitable for driving wheels, gears, and other mechanical parts in miniature robots or automated systems. Its reliability and efficiency make the N20 shaft motor a popular choice among hobbyists and engineers looking to add precise and powerful motion to their projects

#### D. LED

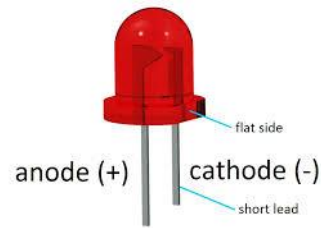

Fig 4. LED

LED stands for light emitting diode. Compared to incandescent light bulbs, LED lighting products provide light up to 90% more effectively. Visible light is produced when an electrical current flows through a microchip, illuminating the small light sources known as LEDs. Minority and majority charge carriers recombine at the junction as current flows through the diode. Photons are released as a result of recombination.

#### E. Blynk IoT

Blynk IoT is an open-source unified IoT platform used to build and maintain smart IoT applications and simplifies the process of creating IoT applications. As a user-friendly interface, it supports a wide range of hardware devices, making it ideal for rapid prototyping and development in IoT projects. It offers features like drag-and-drop widgets and customizable dashboards, It helps users to quickly build and deploy IoT solutions without extensive coding knowledge.

Here when an external device upon successfully authenticated and connected to the WIA-Comm network, users interact with the application through widgets configured in the Blynk IoT platform, the on-off control commands are transmitted to the Blynk Cloud, serve as an intermediary, from which the NodeMCU retrieves the commands and executes the corresponding operations.

## IV. METHODOLOGY

The basic idea of working on the prototype or the model can be briefly described as follows. The configuration of the module for WIA-Comm. system is discussed in this section.

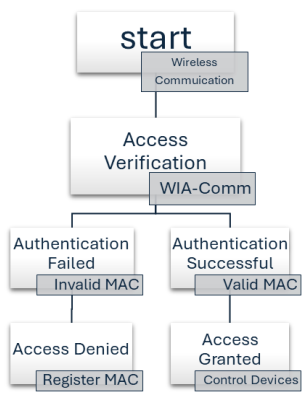

Fig 5. WIA-Comm System Flow diagram

### A. Transmitter part

As we know transmitter is essential in any secured network for ensuring reliable and secure data transmission within the network architecture, we can see the configuration of the WIA-Comm transmitter part as follows. When any external device wants to control the device at the application end, it will raise a request to connect to this secured network. During this stage, the authentication process of the system is initiated at the transmitter part. MAC addresses, also known as Media Access Control addresses, are unique 48-bit hardware identifiers assigned to network interface cards during manufacturing [1]. They serve as the physical address of a network device and play a crucial role in communication capabilities within a network infrastructure. The MAC address or the physical address of the device that is trying to connect is used as an authentication mechanism and decides whether to allow access to the device or not. MAC address is preferred as this address is a static address unique to every device and the MAC address does not change, unlike an IP address. Dynamic IP assignment in a network is a major drawback to using the IP address for authentication purposes, as the IP address is dynamically assigned every time a device is connected to a network.

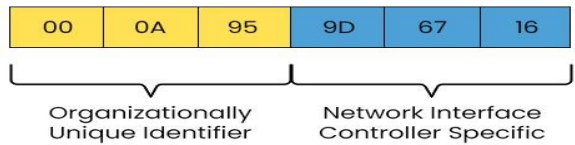

Fig 6. 48-bit MAC address

The MAC address of that device is first retrieved by the ESP8266 module on the control side of the system along with the LoRa module (control board) within its framework. The obtained MAC address is next compared or verified along the predefined list of registered devices' MAC addresses. These registered MAC addresses can be called the list of devices that have access to the system or network. If the MAC address matches or is verified, then the device will be allowed to the network/system. Then the intended signal is passed on to the receiver side of the system which is called the application side which enables control of the devices in the application side. This transfer of control signal is aided using the LoRa modules which have been interfaced for communication purposes.

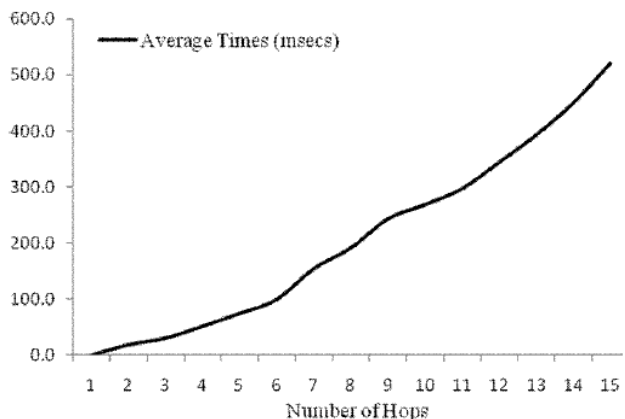

Fig 7. Elapsed time for tracing a Mac address by number of hops [2]

Fig.8. illustrates the transmitter side of the WIA-Comm system, which mandates any external device intending to access the network to install the Blynk application for control purposes. Once passing through the Blynk cloud, the device encounters the security layer of MAC-based authentication, representing a novel scheme not previously utilized in secured networks. Upon successful authentication on the transmitter side, the control signal is transmitted to the receiver side operation LoRa board for execution.

The transmitter side is also known as the control side of the WIA-Comm. System.

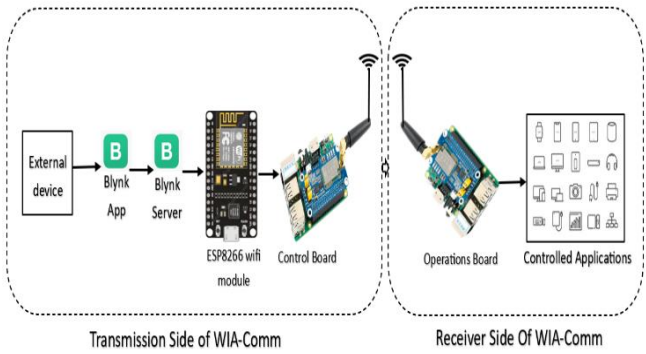

Fig 8. WIA-Comm Trans receiver network

## B. Receiver part

Upon successful authentication, the LoRa receiver board receives execution commands, enabling operations on the application side. This connection extends to an external device requiring secure control. To demonstrate functionality, a breadboard prototype is utilized, comprising an LED and a stepper motor as end-operation components. Upon receiving the command signal, the respective operations are executed on the connected devices.

For demonstration simplicity, remote control of the LED and stepper motor's on-off operations is facilitated via a control template within the Blynk application dashboard. Specifically, switch Widgets tailored for this specific application provide seamless control, allowing users to remotely manage the operations of both the LED and stepper motor.

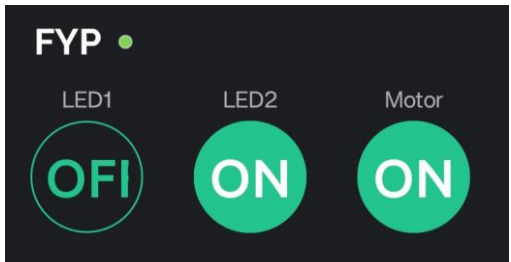

Fig 9. Blynk IoT interface template for WIA-Comm.

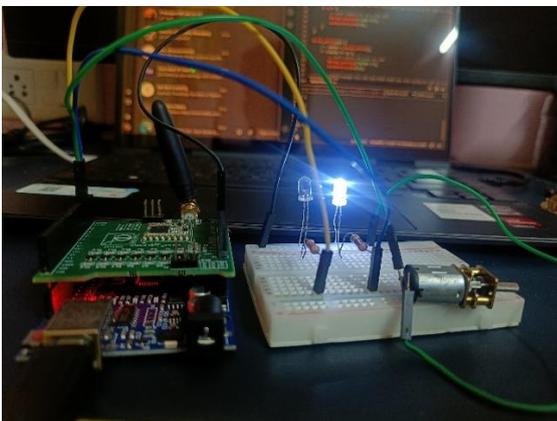

Fig 10. Hardware setup of receiver part of WIA-Comm.

Conversely, if the authentication of the device fails, i.e., the MAC address of the device does not match with the predefined list of MAC addresses, then access to the device to the system or the network is denied. Hence any unknown device cannot illegally enter the system and control the devices on the application side. The receiver side can also be called the application side as it has devices that are controlled through the control side.

By these mechanisms, we can ensure a good amount of security given to the system and prevent any illegal access or control and as a result, can stop any harmful effects or alterations and hence result in smooth and secure functioning of the system.

## V. ADVANTAGES

As we have gone through the details, specifications, and working of the WIA-Comm system, let us analyze and finally look at the advantages or the benefits of having such a system.

- Unauthorized outsiders or intruders are prevented from gaining access to the system and controlling devices or components on the application side.
- The use of LoRa WAN enables Long Range communication which gives provision for remote access and control.
- Blynk's support for diverse hardware enables swift prototyping and deployment of IoT solutions.
- MAC-based authentication simplifies user access control by relying on a device's inherent identifier, eliminating the need for additional credentials or complex authentication mechanisms.
- WIA-Comm provides a clear and traceable means of identifying and authorizing devices accessing the network, enhancing accountability in the event of security breaches or unauthorized activities.

## VI. LIMITATIONS

By nature, anything that has many advantages also comes associated with a few constraints. Yet so many benefits of the system, but there are a few drawbacks that the system possesses which may sometimes be identified as a downside. Some drawbacks may be in the use of components or the conditions and the specifications.

- Given the uniqueness of MAC addresses assigned to individual devices, theft of an authenticated or registered device may result in unauthorized and potentially illegal control of the system by external entities.
- Adding or dropping a device(MAC address) from the list of authorized devices is possible only by the developer himself.
- Damage in one component may lead to improper functioning of the entire system.
- Overloading of one device(eg. A motor) in the peripheral components may lead to the damage of other dependent peripheral components and sometimes also the system.

## VII. RESULTS

The implementation and testing of the WIA-Comm system yielded expected results, showcasing its efficacy in providing secure and authenticated wireless communication for remote device control. Through MAC address-based authentication, LoRa WAN communication, and the Blynk IoT control interface.

A comparative analysis with traditional systems highlighted the system's deployment flexibility for remote locations. Overall, the results underscore the potential of the WIA-Comm system in meeting the growing demand for secure and efficient wireless communication in IoT applications.

Table 1. Comparison table of authentication method

| Authentication Method | Security Level | Ease of Implementation | Vulnerability to Attacks |
|---|---|---|---|
| MAC Address-based Authentication | High | Moderate | Low |
| Password-based Authentication | Moderate | High | Moderate |
| Fingerprint-based Authentication | High | Low | Low |

Table 2. Table of comparative features

| Feature | WIA-Comm System | Traditional Systems |
|---|---|---|
| Authentication | MAC address-based | Password-based |
| Communication | LoRa WAN | Wi-Fi, Bluetooth |
| Control Interface | Blynk IoT | Web interface |
| Security Mechanism | MAC address | Encryption |
| Deployment Flexibility | Remote locations | Local networks |

The result of the serial monitor from the Arduino IDE can be viewed as below:

Fig 9.a. At LoRa transmitter side

Fig 9.b. At LoRa receiver side

## VIII. FUTURE ADD-ONs

As we all know the saying goes, there's always scope for improvement, every product will have some room or scope for improvement or an add-on for better working and functionality of the product. Even in this WIA-Comm. system, there are some areas or functionalities that can be improvised or added so that the effective working of the product can be justified.

- To provide a higher level of security, we can also include a username and password login feature to protect the system in case the device with an authorized MAC address is stolen.
- The Blynk IoT app can be replaced by physical switches when the place of remote control is fixed.
- Also, licensed bands can be used for communication as LoRa supports a wide range of bands for communication. This may lead to improved privacy.
- Also, based on the user requirement, the number of devices that have to be controlled on the application side can be altered, which means more devices like motors, and other sensors can be interfaced to provide more functional coverage.

Table 3. Future Research Directions

| Research Area | Description |
|---|---|
| Enhanced Security | Implementing multi-factor authentication |
| IoT Integration | Integrating with other IoT protocols |
| Scalability | Supporting larger networks and device counts |
| Energy efficiency | Optimizing power consumption for battery-operated devices |